# Emergence of high-temperature superconductivity at the interface of two Mott insulators

Lele Ju[1#], Tianshuang Ren[1#], Zhu Li[2#], Zhongran Liu[2], Chuanyu Shi[1], Yuan Liu[1], Siyuan Hong[1], Jie Wu[3,4], He Tian[2], Yi Zhou[5,6,7*], and Yanwu Xie[1,8*]

[1]Interdisciplinary Center for Quantum Information, State Key Laboratory of Modern Optical Instrumentation, and Zhejiang Province Key Laboratory of Quantum Technology and Device, Department of Physics, Zhejiang University, Hangzhou 310027, China.
[2]Center of Electron Microscope, State Key Laboratory of Silicon Materials, School of Materials Science and Engineering, Zhejiang University, Hangzhou, 310027, China.
[3]Key Laboratory for Quantum Materials of Zhejiang Province, School of Science, Westlake University, Hangzhou 310024, China.
[4]Institute of Natural Sciences, Westlake Institute for Advanced Study, Hangzhou 310024, China
[5]Beijing National Laboratory for Condensed Matter Physics & Institute of Physics, Chinese Academy of Sciences, Beijing 100190, China.
[6]Songshan Lake Materials Laboratory, Dongguan, Guangdong 523808, China.
[7]Kavli Institute for Theoretical Sciences and CAS Center for Excellence in Topological Quantum Computation, University of Chinese Academy of Sciences, Beijing 100190, China.
[8]Collaborative Innovation Center of Advanced Microstructures, Nanjing University, Nanjing 210093, China.
[*]Correspondence to: yizhou@iphy.ac.cn (Y. Z.); ywxie@zju.edu.cn (Y. X.)
[#]These authors contributed equally to this work.

**Abstract:**
Interfacial superconductivity has manifested itself in various types of heterostructures: band insulator-band insulator, band insulator-Mott insulator, and Mott insulator-metal. We report the observation of high-temperature superconductivity (HTS) in a complementary and long expected type of heterostructures, which consists of two Mott insulators, $La_2CuO_4$ (LCO) and $PrBa_2Cu_3O_7$ (PBCO). By carefully controlling oxidization condition and selectively doping $CuO_2$ planes with Fe atoms, which suppress superconductivity, we found that the superconductivity arises at the LCO side and is confined within no more than two unit cells (~2.6 nm) near the interface. A phenomenon of "overcome the Fe barrier" will show up if excess oxygen is present during growth. Some possible mechanisms for the interfacial HTS have been discussed, and we attribute it to the redistribution of oxygen.

## 1. Introduction

Interface plays a central role in cutting-edge science and technology, and can harbor a host of emergent phenomena that could even be absent in corresponding bulk materials. Above all, superconductivity can manifest itself at interfaces involving oxides [1-14]. Typical examples include the emergence of superconductivity at the interfaces of two non-superconducting oxides, such as $LaAlO_3/SrTiO_3$ [2], $La_2CuO_4(LCO)/La_{1.55}Sr_{0.45}CuO_4$ [1], and newly discovered EuO(or $LaAlO_3)/KTaO_3$ [11–13], and the enhancement of superconducting transition temperature ($T_c$) of the single-unit-cell FeSe grown on $SrTiO_3$ [3]. These fascinating observations have been attracting increasing research interests. They not only provide a unique way to create novel superconductors in artificial systems and design state-of-art superconducting devices, but also sheds light on the mechanism of high-temperature superconductivity (HTS), a holy grail in modern physics.

Among all these interfaces, the LCO-based heterostructures [1,4,5,15–19] are of particular interest because LCO is a prototype HTS parent compound. Gozar et al. [1] observed superconductivity in heterostructures consisting of undoped LCO (a Mott insulator) and over-doped (OD) $La_{2-x}Sr_xCuO_4$ (a metal). Similar results were reported in LCO/OD-$La_{2-x}Ba_xCuO_4$ and LCO/OD-$La_{2-x}Ca_xCuO_4$ heterostructures [17,18]. With the help of atomic-scale control of growth and the technique of δ-doping, Logvenov et al. [4] demonstrated that superconductivity occurs within a single $CuO_2$ plane in LCO/(metallic)$La_{1.55}Sr_{0.45}CuO_4$ heterostructures. Wu et al. [15] examined more than 800 different compositions of LCO/$La_{2-x}Sr_xCuO_4$ (0.15 < x < 0.47) heterostructures and found an anomalous doping-level-independence of interfacial superconductivity. While atomically perfect interfaces are of extreme importance and were explored in the aforementioned studies, two very recent experiments [5,20] indicated that the interfacial superconductivity is so robust that the surface roughness is no longer a serious obstruction. As observed by Pavlov et al. [5], superconductivity survives by depositing $Ba_{0.8}Sr_{0.2}TiO_3$ (a band insulator, ferroelectrics) film on LCO singe crystal with 1-2 nm surface roughness. And it was found by Deng et al. [20] that superconductivity exists in LCO/$La_{2-x}Sr_xCuO_4$ heterostructures within a surprisingly wide range of doping, nominally 0.10 ≤ x ≤1.70, even though the $La_{2-x}Sr_xCuO_4$ film has lost its crystalline structure when x ≥ 1.40 [20].

In this work, we report the observation of HTS in heterostructures composed of LCO and $PrBa_2Cu_3O_7$ (PBCO, a Mott insulator). To the best of our knowledge, this is the first demonstration that superconductivity can occur at interfaces of two Mott insulators. This novel type of interfacial superconductor is complementary to existing emergent superconductors at interfaces of band insulator-band insulator [2,11,12], band insulator-Mott insulator [5,21], and Mott insulator-metal [1]. Our experiments revealed that the superconductivity arises in the LCO side and near the PBCO/LCO interface. The thickness of the superconducting layer is of atomic-scale and can be of two LCO unit cells or thinner, which depends on the oxidization condition. Meanwhile, as the first example of HTS, a doped LCO thin film of a few atomic layers is highly desirable on an insulating substrate to reveal the origin of HTS. Thus, our work not only unveils a new category of interfacial superconductors, but also provides a fresh platform to explore the rich physics of HTS.

## 2. Materials and methods

### 2.1. Sample Growth

All the single films and bilayers were deposited on SrLaAlO$_4$ (SLAO)(001) single-crystal substrates by pulsed laser deposition. Targets were prepared by conventional solid-state reaction with nominally stoichiometric compositions. Before growth the substrates were pre-annealed in situ at 800 ºC for 20 minutes, under an atmosphere of $1\times10^{-4}$ mbar molecular oxygen. Three growth conditions were used.

Condition 1: The standard one. The growth temperature was 730 ºC. The laser fluence was ~1 Jcm$^{-2}$. The laser frequence was 4 Hz. The target-substrate distance was ~55 mm. A mixture of 20% ozone and 80% molecular oxygen was used as gas source. During growth a dynamic atmosphere with a flow rate of 3 sccm and a pressure of 0.03 mbar was used. After growth, the samples were post-annealed in situ at 400 ºC for 15 minutes, with a flow rate of 10 sccm and without pumping. After that the samples were cooled to below 60 ºC with the same flow rate and without pumping. Then, all the gas in the chamber was evacuated and instead a 200-mbar molecular oxygen was filled. Using this molecular oxygen atmosphere, the samples were annealed again at ~200 ºC for 30 minutes to balance oxygen content (to remove excess oxygen induced by ozone) and finally cooled to room temperature. The growth was monitored by RHEED, which indicated a layer-by-layer mode (see Fig. S1).

Condition 2: The oxygen-excess one. All the parameters used in this condition are the same as that used in the Condition 1, except for the annealing temperature in the last annealing with 200-mbar molecular oxygen. In this condition a lower annealing temperature of ~150 ºC (vs ~200 ºC) was used.

Condition 3: The ozone-free one. In this condition, ozone was NOT used in all the steps. The growth temperature was 650 ºC. The growth atmosphere was a mixture of 0.077-mbar molecular oxygen and 0.033-mbar nitrous oxide (N$_2$O). After growth all the samples were post-annealed in situ at 500 ºC, in 200-mbar oxygen, for 30 minutes. Other conditions are similar to those of Conditions 1 & 2.

### 2.2. Electrical contacts and transport measurements

The contacts were made with Ag film electrodes deposited on the surface of the samples by electron-beam evaporation. These electrodes were further connected for measurements using Al wires by an ultrasonic wire bonding machine. A DC measurement method was used for the $R_{sq}(T)$ measurements in a commercial cryostat.

### 2.3. Scanning transmission electron microscopy (STEM) and energy-dispersive x-ray spectroscopy (EDS) mapping.

Cross-sectional specimens for electron microscopy investigations were prepared by a FEI quanta 3D FEG Focused Ion Beam. Atomic resolution HAADF-STEM images and EDS mappings were acquired by a spherical aberration corrected electron microscope equipped with 4 Super-X EDS detectors (FEI Titan G2 80-200 ChemiSTEM).

### 2.4. *X-ray diffraction (XRD)*

The XRD data were taken using a monochromated Cu-K$_\alpha$ source on a 3-kW high-resolution Rigaku Smartlab system.

### 3. Results and discussion

#### 3.1. *Structural characterizations of PBCO/LCO heterostructure*

The standard growth condition (Condition 1) was used if not specified. Atomic force microscopy measurements show that the sample surfaces are smooth (Fig. S2) and the surface roughness of a typical PBCO(6 uc)/LCO (10 uc) heterostructure is ~0.47 nm, smaller than one unit cell of PBCO (~1.1 nm) or LCO (~1.3 nm). X-ray diffraction data reveal that the growth is epitaxial (Figs. 1(a) and S3). Scanning transmission electron microscopy (STEM) high-angle annular dark-field (HAADF) images (Figs. 1(b) and 1(c)) show that the interface is abrupt. The energy-dispersive x-ray spectroscopy (EDS) map (Fig. 1(d)) and line scan (Fig. S4) show that the interface is chemically abrupt, and the atomic intermixing is limited to about 1 uc.

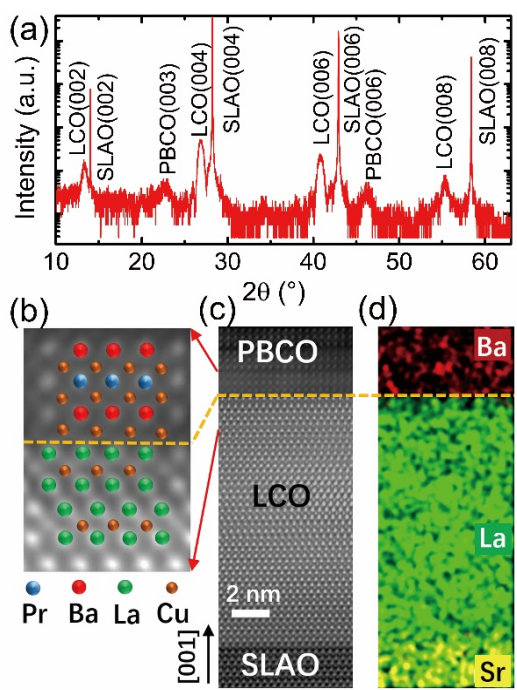

**Fig. 1.** Structural characterizations. (a) X-ray diffraction of a PBCO(6 uc)/LCO(10 uc) heterostructure. For the notation of heterostructures, the right component always denotes the one next to the SLAO substrate. (b, c)HAADF-STEM images and (d) EDS elemental mapping of a representative PBCO/LCO heterostructure. The dashed line indicates the position of the PBCO/LCO interface. The interface is abrupt and intermixing is not significant.

#### 3.2. *Emergence of superconductivity in LCO layer*

We performed transport measurements on single-phase films as well as heterostructures. Figure 2(a) shows that both the PBCO and LCO single-phase films are insulating. In contrast, as shown in Fig. 2(b), the PBCO/LCO heterostructure (the thick black line) exhibits a metallic behavior in high-temperature range and turns into superconductivity at low temperatures. The superconductivity onset temperature is ~30 K, and the zero-resistance occurs at ~19 K. The superconductivity can be fully suppressed if 1-uc SLAO layer is inserted between PBCO and LCO (Fig. 2(b)), implying that it comes from the PBCO/LCO interface. In order to investigate the interfacial superconductivity further, we replaced a fraction of Cu atoms in PBCO or LCO by Fe atoms, which was realized by depositing the corresponding films with Fe-doped targets. The presence of Fe can suppress superconductivity in copper oxides efficiently (see Fig. S5) [22,23]. So that different patterns of Fe-doping give rise to distinct transport features. As shown in Fig. 2(b), the Fe-doping in the LCO side leads to an insulating state and the $R_{sq}(T)$ is much larger than that of PBCO/LCO; in sharp contrast, the Fe-doping in the PBCO side remains a superconducting state and the $R_{sq}(T)$ is very close to that of PBCO/LCO. Thus, we conclude that the observed superconductivity and the overall conductance is in the LCO layer near the interface.

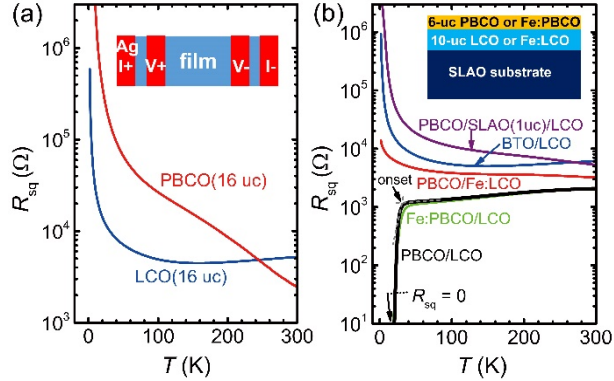

**Fig. 2.** Temperature-dependent squared resistance, $R_{sq}$. (a) PBCO and LCO single-phase films grown on SLAO. (b) PBCO/LCO heterostructures (with or without Fe doping. Fe:LCO and Fe:PBCO stand for $La_2Cu_{0.95}Fe_{0.05}O_4$ and $PrBa_2Cu_{2.95}Fe_{0.05}O_7$, respectively.) grown on SLAO. Inset of (a): a schematic view of the measurement configuration. Inset of (b): a schematic cross view of heterostructures. For comparison, a PBCO(6 uc)/SLAO(1 uc)/LCO heterostructure, in which 1-uc SLAO layer was inserted between LCO and PBCO, and a heterostructure composed of 6.6-nm $BaTiO_3$ (BTO) and 10-uc LCO are shown in (b).

### 3.3. Superconducting layer confined at the interface

The RHEED-assisted layer-by-layer growth method allows us to dope a designated 1-uc-layer of LCO with Fe, so called "δ-doping" (here in a reverse meaning since the doped layer will not be superconducting). This technique was used previously by Logvenov et al. [4] to locate the single superconducting $CuO_2$ plane in $LCO/La_{1.55}Sr_{0.45}CuO_4$ heterostructures precisely. Here we utilized this technique to identify the thickness and the position of the superconducting layer in LCO. As illustrated in Fig. 3, the superconductivity will be fully suppressed if the N = -1 unit cell of LCO (the one closest to the interface) is replaced by Fe:LCO, i.e., $La_2Cu_{0.95}Fe_{0.05}O_4$. In contrast, if the N = -2 unit cell is doped by Fe, the superconductivity will survive, although its $T_c$ is somewhat lower than that of the undoped one (the dotted line in Figs. 3(b) and 3(c)); if N = -3 unit cell is doped, the temperature-dependence of resistance $R(T)$ will almost coincide with the undoped one. This coincidence suggests that the conducting and superconducting channel is dominated by charge carriers in LCO that are confined within one or two unit cells next to the interface.

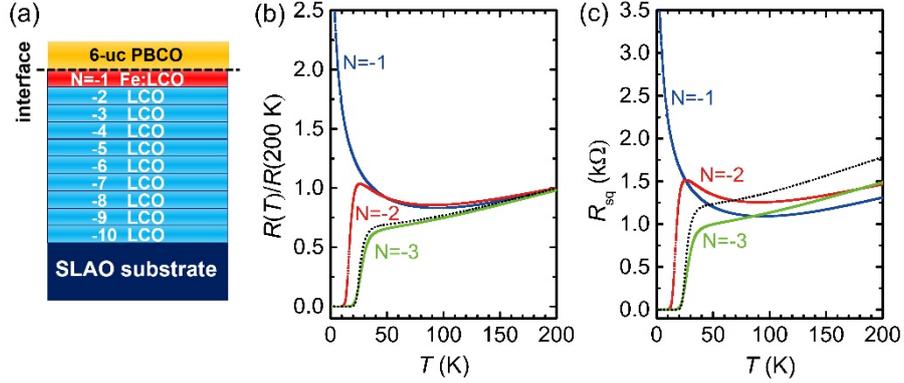

**Fig. 3.** Replacing one unit cell of LCO with Fe:LCO in PBCO(6 uc)/LCO(10 uc) heterostructures. (a) A sketch of the structure. The positions of LCO unit cells are as labeled. Temperature-dependent (b) normalized resistances and (c) the corresponding squared resistances of δ-doped PBCO/LCO heterostructures. The dotted line is the PBCO/LCO heterostructure shown in Fig. 2(b) (the thick line) for comparison.

Note that the high substrate temperature during growth may cause some Fe atoms diffuse from the Fe:LCO unit cell to neighboring LCO unit cells. Thus the "actually clean" LCO layer near the interface should be thinner than 1 uc for N = -2 and 2 uc for N = -3. With this consideration in mind, we indeed set the upper limit of the superconducting layer thickness, $d_{sc} \leq 2$ uc. The $d_{sc}$ was also estimated independently by Hall effect measurements. The data measured at 50 K on the PBCO(6 uc)/LCO(10 uc) heterostructure result in a hole density of ~$5.8 \times 10^{14}$ cm$^{-2}$ at the interface (Fig. S6). Taking a bulk carrier density of $2.1 \times 10^{21}$ cm$^{-3}$ from the optimally doped $La_{2-x}Sr_xCuO_4$ [24] as a reference, we estimated $d_{sc}$ to be ~2.8 nm, which is close to the thickness of 2-uc LCO.

Furthermore, we gradually reduced the thickness of LCO from 10 to 1 uc. It turns out that the temperature-dependent $R_{sq}(T)$ will be quite similar when the thickness of LCO exceeds or equals to 2 uc (Fig. S7). This similarity is strongly indicative of the fact that charge carriers are confined to the interface.

### 3.4. *Enhanced superconductivity*

In addition to the above-discussed samples prepared under the standard growth condition (Condition 1), we also fabricated PBCO/LCO heterostructures under a more oxygen-excess growth condition (Condition 2). First of all, we would like to emphasize that this new condition is only moderately oxygen-excess and will not by itself drive an insulating LCO film into a superconducting one [1,25]. This is evidenced by the fact that both PBCO and LCO single-phase films grown under Condition 2 remain highly insulating, as illustrated in Fig. 4(a). In the meantime, Fig. 4(b) (black line) shows that the undoped PBCO/LCO heterostructure grown under Condition 2 exhibits enhanced superconductivity with a higher onset $T_c$ ~36 K. This enhancement can be attributed to the increase of hole concentration due to the excess oxygen.

Even more strikingly, the enhanced interfacial superconductivity can overcome the Fe barrier and move toward the bulk, which is in sharp contrast to the ones grown under the standard condition. To reveal this marvelous phenomenon, we prepared samples with the PBCO(6 uc)/Fe:LCO(n uc)/LCO[(10-n) uc] heterostructures, where the unit cell number of Fe:LCO layer, n, ranges from 1 to 5. As shown in Fig. 4(b), when n=1, the N=-1 Fe δ-doped PBCO/LCO heterostructure exhibits a comparable superconductivity with the undoped one. Even when n=5, the topmost five LCO unit cells near the interface (N=from -1 to -5) are all Fe-doped, namely, the middle layer is Fe:LCO(5 uc), the heterostructure still remains superconducting. However, the sample will be highly insulating if the PBCO layer is absent (Fig. 4(b), the grey dashed line), revealing that the presence of PBCO/LCO(Fe:LCO) interface is crucial to the superconductivity.

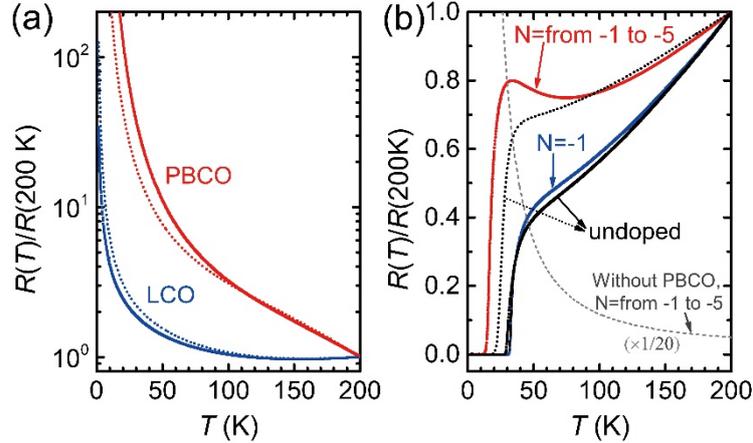

**Fig. 4.** Films and heterostructures grown under oxygen-excess condition. (a) PBCO and LCO single-phase films grown on SLAO. The dotted lines are the data shown in Fig. 2(a) for comparison. (b) PBCO(6 uc)/LCO (10 uc) and PBCO(6 uc)/Fe:LCO(n uc)/LCO[(10-n) uc] heterostructures grown on SLAO. The positions of Fe-doped LCO unit cells are denoted using the same labels as schemed in Fig. 3(a). The dotted line is the PBCO/LCO heterostructure shown in Fig. 2(b) for comparison. The grey dashed line illustrates a sample without the top PBCO layer, that is, Fe:LCO(5 uc)/LCO(5 uc), for comparison.

### 3.5. The origin of the superconductivity

We now discuss the possible mechanism for the occurrence of HTS in PBCO/LCO heterostructures. It is generally accepted that the physics of cuprate HTS is doping a Mott insulator [26]. Comprehensive studies have suggested that hole doping can be achieved by oxygen excess, cation substitution, or charge transfer. Let us examine all these possibilities as follows.

First, the possibility of ozone-induced oxygen excess can be excluded because the LCO single-phase films are always insulating although they are grown under the same conditions as the heterostructures (Figs. 2 and 4). To further exclude this possibility, we have also grown samples in a completely ozone-free atmosphere (Condition 3), and observed superconductivity in PBCO/LCO heterostructures that share essentially the same $T_c$ and $R$(T) as the samples grown under the standard condition (Condition 1) (Fig. S8). Thus, we infer that the HTS is not caused by the ozone-induced oxygen excess and instead, the presence of PBCO is crucial for its occurrence.

The second possibility is the cation substitution, which might arise from the diffusion of Ba from PBCO into LCO. However, STEM and EDS measurements (Figs. 1(d) and S4) show that this diffusion channel is not significant and the diffusion length is less than 2 nm. Meanwhile, in the aforesaid phenomenon of "overcome the Fe barrier" illustrated in Fig. 4(b), for the heterostructure PBCO(5 uc)/Fe:LCO(5 uc)/LCO(5 uc) grown under Condition 2, the HTS was detected in the LCO region that is more than 6 nm away from the PBCO/Fe:LCO interface and is out of the range of Ba diffusion. Therefore the Ba-diffusion mechanism can be ruled out in the PBCO/LCO heterostructures grown under Condition 2. Note that there is only small difference between the two growth conditions, Condition 1 and 2 (the difference is that the final annealing temperature in molecular oxygen was decreased from ~200 to ~150 ºC; see Materials and Methods), and the key findings are essentially the same (Figs. 2 and 4). Thus, it is natural to assume that the heterostructures grown under these two conditions (Condition 1 and Condition 2) share the same HTS mechanism. As an independent evidence to reject the Ba diffusion mechanism, we grew BaTiO$_3$(BTO)/LCO heterostructures under the standard condition and found no superconductivity (Fig. 2(b)), which further supports that the Ba diffusion is not the dominant mechanism for the observed HTS in PBCO/LCO.

The third possibility is the charge transfer, *i.e.*, the redistribution of holes between PBCO and LCO. Charge transfer was proposed to be the dominant mechanism for the HTS in LCO/La$_{1.55}$Sr$_{0.45}$CuO$_4$ heterostructures [1,27]. Nevertheless, there is a huge difference between La$_{1.55}$Sr$_{0.45}$CuO$_4$ and PBCO: the former is a metal with a high hole concentration, while the latter is an insulator. Previous studies suggested that PBCO is not a good reservoir of holes, even though its role on charge transfer is still controversial. For instance, YBa$_2$Cu$_3$O$_7$(YBCO)/PBCO supperlattices [28] and YBCO layers sandwiched by PBCO [29] displayed enhanced HTS in ultrathin YBCO, but the capping PBCO was generally regarded as a protecting layer. While Terashima *et al.* [29] advocated that the PBCO layer provides holes to the YBCO layer, Affronte *et al.* [30] suggested no significant charge transfer, and Cieplak *et al.* [31] even argued a reverse charge transfer. On the other hand, if the phenomenon of "overcome the Fe barrier" was explained by charge transfer, the amount of holes that are needed to be transferred from PBCO to LCO would be too large (> $10^{15}$ cm$^{-2}$) to be likely, since it would cause an unrealistically huge built-in electrical field at the interface. Thus, we argue that charge transfer is not the dominant mechanism for the observed HTS.

We now turn to the remaining possibility, the oxygen non-stoichiometry induced by the presence of PBCO/LCO interface, in which oxygen atoms transfer from PBCO to LCO and provide holes. A crucial difference between charge transfer and oxygen transfer is that the issue of huge built-in electrical field will be circumvented in the latter case, even though a sufficient amount of induced holes has been taken into account. Since LCO itself is a reservoir of oxygen, the oxygen transfer will modify the charge background in a gradual way. The oxygen transfer can also give rise to the phenomenon of "overcome the Fe barrier". Note that a similar oxygen transfer scenario was indicated in a previous study on electron-doped cuprate heterostructures composed of $La_{2-x}Ce_xCuO_4$ and $Pr_{2-x}Ce_xCuO_4$ superlattices [32]. Due to the technical difficulty, at present we cannot directly detect the diffusion of oxygen, unlike the diffusion of cations, with the local STEM and EDS probes. Aside from this difficulty our key findings can be consistently explained by this scenario. Finally, the extreme sensitivity of HTS with oxidization conditions during the growths of heterostructures (Figs. 2 & 4) supports that oxygen plays a crucial role in the observed phenomena.

## 4. Summary and outlook

We have demonstrated that a robust HTS can be created in heterostructures of two Mott insulators, PBCO and LCO. The superconductivity can be manipulated by modifying the oxidization condition during the growth of the heterostructures. Have in mind that both PBCO and LCO are insulating, the conductance is dominated by the superconducting channel, which is ideal for studying the intrinsic transport properties of HTS copper oxides. Moreover, the present observation provides a practical way to make an ultrathin (2 uc or even thinner) superconducting LCO layer that is of potential importance in studying the HTS mechanisms and in device applications.

**Conflict of interest**
The authors declare no competing interests.

**Acknowledgments**
We thank X. H. Chen, G. H. Cao, Z. A. Xu, K. Jin, H. Yao, F. C. Zhang, and H. Y. Hwang for useful discussions. This work was supported by the National Key Research and Development Program of China (2017YFA0303002), National Natural Science Foundation of China (11934016, 12074334, 12034004 and 11774306), the Key Research and Development Program of Zhejiang Province, China (2020C01019, 2021C01002), and Natural Science Foundation of Zhejiang Province, China (Grant No. LD21E020002). Y. Z. is also supported by the Strategic Priority Research Program of Chinese Academy of Sciences (No. XDB28000000).

**Author contributions**
Y. X. and Y. Z. conceived the idea and designed the experiments. L. J., T. R., and C. S.

fabricated the samples and characterized them with AFM and XRD. L. J., T. R., Y. L. and S. H. made the transport measurements. Z. Li, Z. L., and H. T. measured the STEM data. J. W contributed to the analysis of the results. Y. X. and Y. Z wrote the manuscript with contributions from all authors.

**Additional information**
**Supplementary Materials** is available for this paper.

# Supplemental Materials for

# Emergence of high-temperature superconductivity at the interface of two Mott insulators

Lele Ju, Tianshuang Ren, Zhu Li, Zhongran Liu, Chuanyu Shi, Yuan Liu, Siyuan Hong, Jie Wu, He Tian, Yi Zhou[*], and Yanwu Xie[*]

Correspondence to: yizhou@iphy.ac.cn (Y. Z.); ywxie@zju.edu.cn (Y. X.)

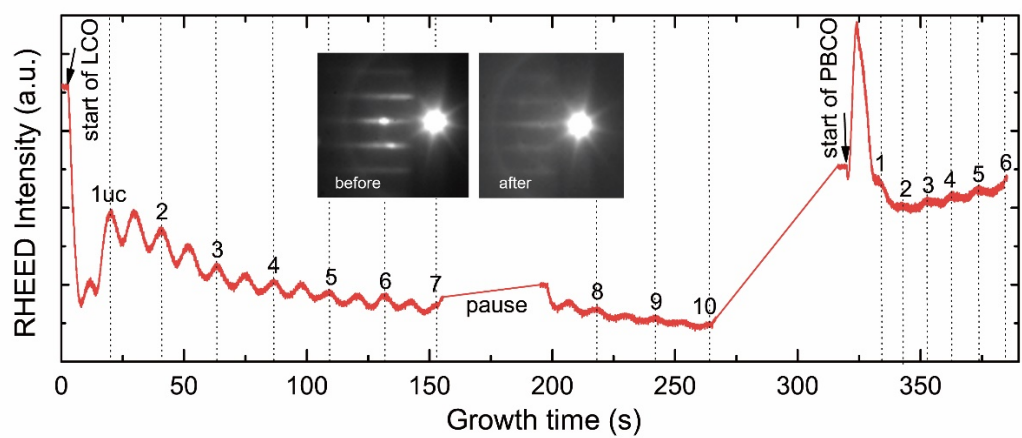

**Fig. S1.** Oscillations of RHEED intensity during the growth of a heterostructure of 10-uc LCO and 6-uc PBCO. Note that one unit cell (uc) LCO (PBCO) corresponds to two (one) periods of oscillations. Insets: RHEED patterns before and after the growth.

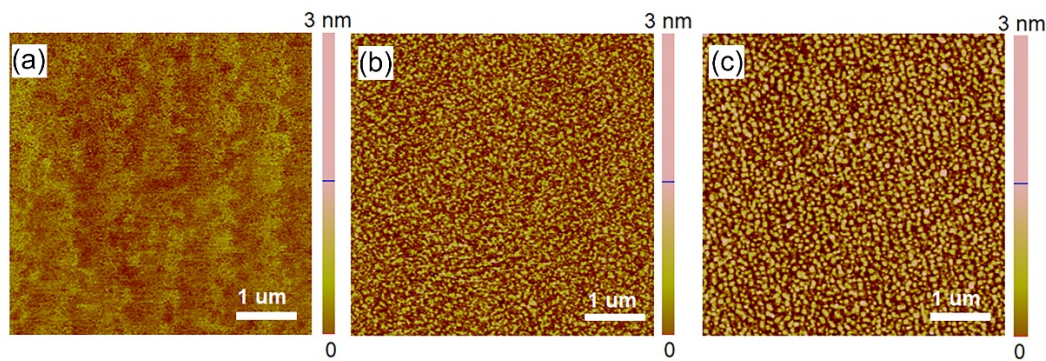

**Fig. S2.** Atomic force microscopy (AFM) images for (a) a 16-uc (21-nm) LCO single film, (b) a PBCO(6 uc)/LCO(10 uc) heterostructure, and (c) a 16-uc (18-nm) PBCO single film grown on SLAO substrates. The root mean square roughness over the whole 5 μm by 5 μm area for (a), (b), and (c) is 0.21, 0.47, and 0.68 nm, respectively.

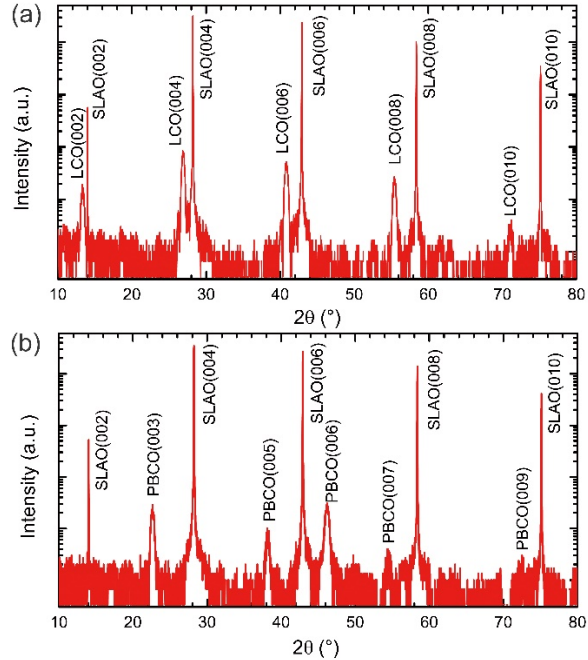

**Fig. S3.** Representative θ-2θ x-ray diffraction (XRD) scans for (a) a 16-uc (21-nm) LCO and (b) a 16-uc (18-nm) PBCO single-phase films grown on SLAO(001) substrates. These data indicate that both kinds of films are highly crystalline and epitaxially grown.

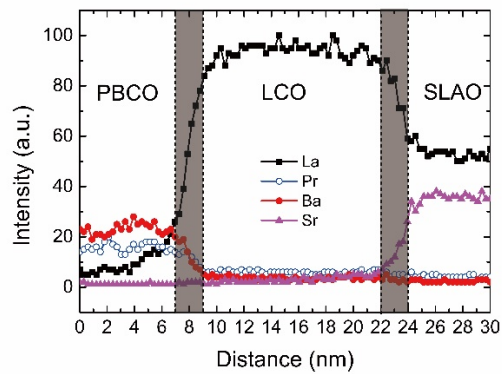

**Fig. S4.** EDS elemental line scans across a PBCO(6 uc)/LCO(10 uc) sample. The shadowed areas highlight the PBCO/LCO and LCO/SLAO interfaces. The EDS intensities decay in ~2 nm. These data indicate that there are no significant cation intermixing at the interfaces. The diffusion of Ba into LCO should be well below 2 nm, probably one unit cell, which sets an upper limit.

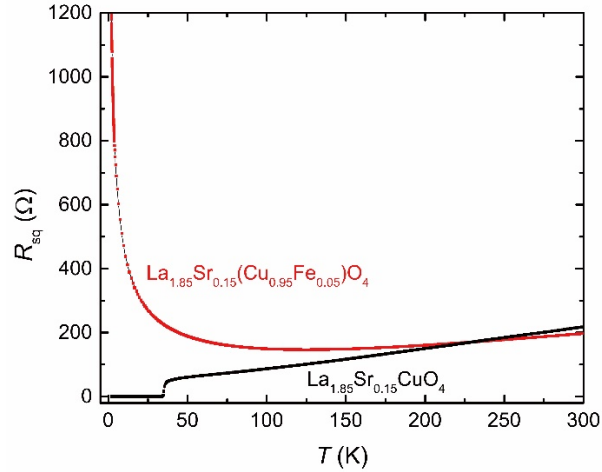

**Fig. S5.** Temperature-dependent $R_{sq}$ of 16-uc undoped and Fe-doped $La_{1.85}Sr_{0.15}CuO_4$ films grown on SLAO. Superconductivity can be efficiently suppressed by partially replacing Cu with Fe.

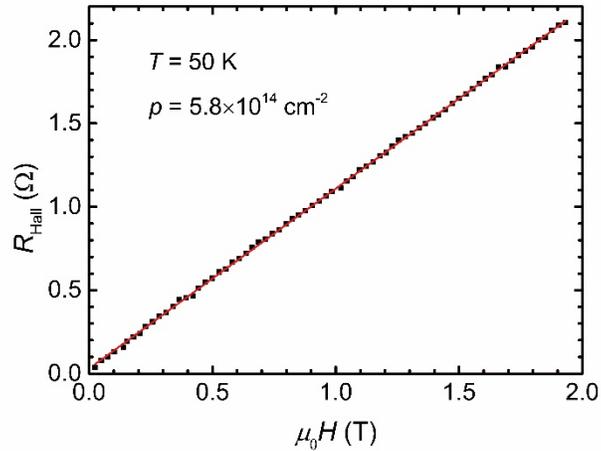

**Fig. S6.** Dependence of $R_{Hall}$ on magnetic field $\mu_0H$ for the PBCO(6 uc)/LCO(10 uc) heterostructure grown under Condition 1. This sample was patterned into a 10 μm-wide Hall-bar configuration by conventional photolithography and lift-off techniques.

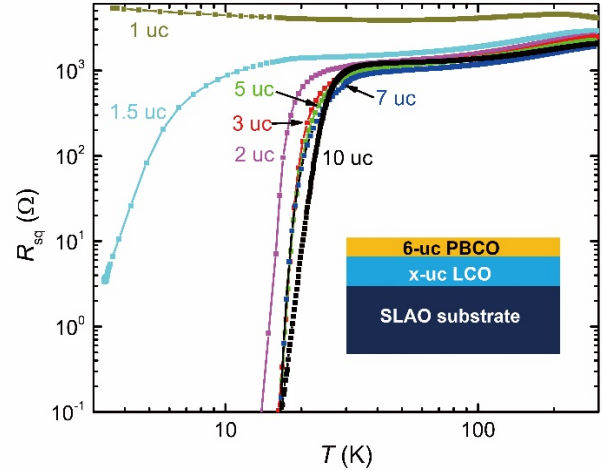

**Fig. S7.** Temperature-dependent $R_{sq}$ for PBCO/LCO heterostructures of different LCO thicknesses. The heterostructures were fabricated under the standard growth condition (Condition 1).

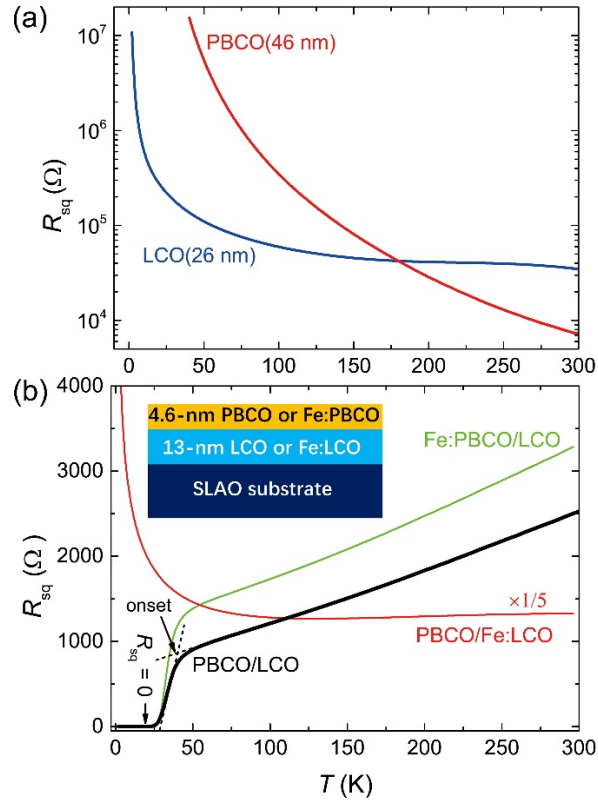

**Fig. S8.** Samples grown without ozone (Condition 3). (a) PBCO and LCO single-phase films grown on SLAO. (b) PBCO/LCO heterostructures (with or without Fe doping. Fe:LCO and Fe:PBCO stand for $La_2Cu_{0.95}Fe_{0.05}O_4$ and $PrBa_2Cu_{2.95}Fe_{0.05}O_7$, respectively.) grown on SLAO. Inset: A schematic cross view of heterostructures. These data demonstrate that the observed superconductivity in PBCO/LCO heterostructures is intrinsic, and is not caused by the ozone-induced oxygen excess in LCO.